\documentclass[twocolumn,floatfix,superscriptaddress,showpacs,showkeys]{revtex4}
\usepackage{graphicx}
\usepackage{amsmath}
\usepackage{booktabs}
\usepackage{amssymb}
\usepackage{multirow}
\usepackage{makecell}
\usepackage{textcomp}
\usepackage{subfigure}
\usepackage{phonetic}
\usepackage{extarrows}
\usepackage{color}
\usepackage{float}
\usepackage{slashed}
\usepackage[colorlinks,citecolor=blue,linkcolor=blue]{hyperref}
\usepackage[utf8]{inputenc}

\begin{document}

\title{Pseudoscalar Yukawa Preheating via Effective Action Approach} 

\author{Xi-Bin Li}
\email{lxbimnu@imnu.edu.cn}
\affiliation{College of Physics and Electronic Information, Inner Mongolia Normal University, 81 Zhaowuda Road, Hohhot, 010022, Inner Mongolia, China}
\affiliation{Inner Mongolia Key Laboratory of Applied Condensed Matter Physics, Inner Mongolia Normal University, 81 Zhaowuda Road, Hohhot, 010022, Inner Mongolia, China}
\author{De-Tian Zhao}
\affiliation{College of Physics and Electronic Information, Inner Mongolia Normal University, 81 Zhaowuda Road, Hohhot, 010022, Inner Mongolia, China}
\affiliation{Inner Mongolia Key Laboratory of Applied Condensed Matter Physics, Inner Mongolia Normal University, 81 Zhaowuda Road, Hohhot, 010022, Inner Mongolia, China}
\author{Ku Fu}
\affiliation{College of Physics and Electronic Information, Inner Mongolia Normal University, 81 Zhaowuda Road, Hohhot, 010022, Inner Mongolia, China}
\affiliation{Inner Mongolia Key Laboratory of Applied Condensed Matter Physics, Inner Mongolia Normal University, 81 Zhaowuda Road, Hohhot, 010022, Inner Mongolia, China}
\author{Jun-Yang Jiao}
\affiliation{College of Physics and Electronic Information, Inner Mongolia Normal University, 81 Zhaowuda Road, Hohhot, 010022, Inner Mongolia, China}
\affiliation{Inner Mongolia Key Laboratory of Applied Condensed Matter Physics, Inner Mongolia Normal University, 81 Zhaowuda Road, Hohhot, 010022, Inner Mongolia, China}

\begin{abstract}
In this work, we investigate helicity-dependent fermion production driven by a pseudoscalar Yukawa interaction within the one-loop effective action framework. 
By integrating out the fermionic fields and employing the Schwinger proper time representation, we derive an analytical expression for the fermion pair production probability near each zero crossing of the oscillating inflaton field. 
We further introduce a helicity transfer matrix to describe the evolution of fermionic modes through successive non-adiabatic production events. 
For an undamped inflaton background, the temporal symmetry around successive zero crossings leads to the cancellation of helicity-mixing contributions after an even number of events, preventing the accumulation of helicity asymmetry. 
We then extend the analysis to a damped oscillatory background, in which the decreasing inflaton amplitude breaks this temporal symmetry and modifies the transfer matrix at each crossing. 
The resulting non-commuting transfer matrices produce a persistent helical spectrum, as demonstrated by the probabilities as a function of momentum obtained after successive scattering events. 
A comparison with scalar Yukawa preheating further confirms the role of helicity mixing in generating the helical asymmetry induced by pseudoscalar mechanism. 
These results establish an effective action framework for studying helicity asymmetry during fermionic preheating and provide insight into the influence of the inflaton's dynamical evolution on particle production.
\end{abstract}
\maketitle

\section{Introduction} \label{sec_intro}

In the standard cosmological paradigm, a transitional epoch is believed to intervene between the end of inflation and the onset of the radiation-dominated era. 
During this stage the energy density stored in the inflaton condensate is converted into radiation. 
Depending on whether the conversion proceeds perturbatively or non-perturbatively, the epoch is referred to as reheating \cite{reheating1} or preheating \cite{preheating_boson01,preheating1}, respectively. 
Bosonic preheating \cite{preheating_boson02,preheating_boson1,preheating_boson2,preheating_boson3,preheating_boson4} is characterised by highly efficient, explosive particle production driven by the coherent oscillations of the inflaton, 
allowing substantial occupation numbers even when single-particle decay is kinematically forbidden. 
An effective-action approach to bosonic preheating has recently been developed \cite{effective_boson_preheating}. 
It has been recognised \cite{preheating_fermion1,preheating_fermion2,preheating_fermion3,preheating_fermion4,preheating_fermion5,preheating_fermion6,preheating_fermion7,preheating_fermion8} that fermionic preheating can likewise be efficient, 
despite the limitations imposed by Pauli blocking. 
In particular, preheating mediated by a pseudoscalar (axion-like) Yukawa interaction is predicted to generate a net helical asymmetry among the produced fermions \cite{Yukawa_preheating,Yukawa_preheating1}. 
Virtually all existing scenarios assume high-scale preheating and therefore predict the existence of new particles heavier than those of the Standard Model. More recently, 
low-scale QCD preheating has been proposed \cite{QCD1,QCD2}, featuring a dynamical, non-adiabatic baryon chemical potential.

Although direct observation of reheating and preheating remains challenging, advances in observational techniques are providing new opportunities to investigate this critical and phenomenologically rich epoch of the early universe. 
Certain physical signatures generated during reheating may survive thermalization or remain partially decoupled from the thermal plasma, thereby preserving valuable information about the underlying processes. 
Among these potential observational probes, gravitational waves have attracted considerable attention, as their primordial spectra may encode information about the energy scale, duration, 
and dynamical evolution of reheating, as well as particle-production processes occurring during this epoch \cite{preheating_GW1,preheating_GW2,preheating_GW3,preheating_GW4,preheating_GW5,preheating_GW6,preheating_GW7,preheating_GW8,preheating_GW9,preheating_GW10,preheating_GW11,preheating_GW12,preheating_GW13,preheating_GW14,preheating_GW15,preheating_GW16}. 
Primordial non-Gaussianity provides another possible avenue for probing reheating, since nonlinear interactions and the coupling between different fields can leave characteristic signatures in the primordial perturbation statistics \cite{preheating_NG1,preheating_NG2,preheating_NG3,preheating_NG4,preheating_NG5,preheating_NG6,preheating_NG7,preheating_NG8,preheating_NG9,preheating_NG10,preheating_NG11,preheating_NG12}. 
In addition, the non-equilibrium dynamics of the early universe may contribute to the generation and evolution of primordial magnetic fields, whose properties could retain information about the underlying physical processes \cite{preheating_PB1,preheating_PB2,preheating_PB3,preheating_PB4,preheating_PB5,preheating_PB6,preheating_PB7,preheating_PB8}. 
Reheating-related fluctuations and nonlinear phenomena may also lead to the formation of primordial black holes, offering potential constraints on the primordial density perturbations and the evolution of the early universe \cite{preheating_BH1,preheating_BH2,preheating_BH3,preheating_BH4,preheating_BH5,preheating_BH6,preheating_BH7,preheating_BH8,preheating_BH9}. 
Furthermore, symmetry-breaking processes and phase transitions associated with the early universe may produce topological defects, such as cosmic strings and domain walls, which could provide additional observational signatures \cite{preheating_TD1,preheating_TD2,preheating_TD3,preheating_TD4}. 
Collectively, these signals offer complementary and potentially powerful indirect probes of the preheating and reheating epoches, helping to constrain the dynamics of the post-inflationary universe.


Yukawa interactions between oscillating scalar fields and fermions constitute a well-established mechanism for fermion production in the post-inflationary universe \cite{Yukawa_preheating,Yukawa_preheating1,preheating_fermion1}. 
In certain preheating models the produced fermions can acquire masses substantially larger than that of the inflaton itself \cite{preheating_fermion5}. 
This feature is of particular relevance for leptogenesis scenarios in which the observed baryon asymmetry is generated by the non-equilibrium decays of heavy right-handed neutrinos 
whose masses lie near the Grand Unified Theory scale \cite{GUT1,GUT2,GUT3}. 
Preheating also plays a crucial transitional role in the cosmological dynamics of rotating axion-like particles: it determines the moment at which the axion 
evolves from an initial rotational phase (capable of supporting baryogenesis) into the coherent oscillatory regime that yields cold dark matter \cite{bary1}. 
In parallel, the gravitational production of massive fermions during inflation and their subsequent cosmological evolution have been extensively studied \cite{WKB1,WKB2,WKB3}.

In this work, we investigate fermionic preheating driven by a pseudoscalar Yukawa interaction within the one-loop effective-action framework. 
By integrating out the fermionic fields and employing the Schwinger proper time representation, 
we first derive the effective action and obtain an analytical expression for the fermion pair-production probability associated with an individual non-adiabatic crossing. 
We then introduce a transfer-matrix description in the helical basis to characterize the evolution of fermionic modes through successive production events. 
The undamped oscillatory background is examined first to identify the role of temporal symmetry in the cancellation of helicity-mixing contributions. 
We subsequently extend the analysis to a damped inflaton background, for which the transfer matrices associated with different production events become dependent on the crossing index and must be combined through an ordered product. 
The resulting momentum-dependent production probabilities are evaluated for different numbers of successive events. 
Finally, we compare the pseudoscalar interaction with the conventional scalar Yukawa coupling and discuss the connection between the effective-action description and the Bogoliubov coefficient approach. 
Our results highlight the combined importance of the interaction structure and the dynamical evolution of the inflaton background in determining the helicity composition of fermions produced during preheating.

The remainder of this paper is organized as follows. In Sec. \ref{sec_preheating}, we derive the one-loop effective action for fermions coupled to the inflaton through a pseudoscalar Yukawa interaction and 
obtain the particle-production probability associated with a single zero crossing of the inflaton field. 
In Sec. \ref{sec_heli_asymmetry}, we develop the transfer-matrix formalism in the helical basis and investigate the accumulation of fermionic production amplitudes over successive zero crossings in an undamped oscillatory background. 
In Sec. \ref{sec_extension}, we extend the analysis in three directions: we first compare the pseudoscalar and scalar Yukawa interactions, 
then examine the effects of inflaton damping on the evolution of the transfer matrices and the resulting fermionic spectra, 
and finally discuss the connection between the effective-action formalism and the conventional Bogoliubov-coefficient approach. 
Section \ref{conclusion} summarizes our main results and presents concluding remarks.

\section{One-loop effective action for Pseudoscalar Yukawa preheating \label{sec_preheating}}
\subsection{The model}

A real scalar inflaton with mass $m_\phi$ is assumed to drive cosmic inflation in the early universe and subsequently couples to Standard Model fields. 
In the present work, we study a preheating scenario in which this inflaton interacts with Dirac fermions $\psi$ of constant mass $m_\psi$ through a trilinear pseudoscalar (axion-like) Yukawa coupling of strength $g$. 
The corresponding action is
\begin{align}
	S[\phi,\bar\psi,\psi]=S_0[\phi]+S_0[\bar\psi,\psi]+S_Y[\phi,\bar\psi,\psi], \label{S_total}
\end{align}
where the scalar inflaton, Dirac fields, and the pseudoscalar Yukawa interaction contributions, respectively, read \cite{pseudoscalar1,pseudoscalar2}
\begin{subequations}  \label{actions}
\begin{align}
	&S_{0}[\phi]=\int\mathrm{d}^{4}x\left(\frac{1}{2}\partial_{\mu}\phi\,\partial^{\mu}\phi - \frac{1}{2}m_{\phi}^{2}\phi^{2}\right), \label{s_phi} \\
    &S_{0}[\bar{\psi},\psi]=\int\mathrm{d}^{4}x\,\bar{\psi}\left(\mathrm{i}\gamma^{\mu}\partial_{\mu}-m_{\psi}\right)\psi, \label{S_psi} \\
    &S_{Y}[\phi,\bar{\psi},\psi]=-\int\mathrm{d}^{4}x\,g\phi\,\bar{\psi}\, \mathrm{i}\gamma^{5}\psi. \label{S_Y}
\end{align} \end{subequations}
Here $\phi$ denotes the inflaton, while $\psi$ and $\bar\psi=\psi^\dagger\gamma^0$ are the Dirac spinor and its adjoint. 
The interaction term $-g\phi\bar\psi\mathrm{i}\gamma^5\psi$ realises the axion-fermion coupling. We adopt the Dirac representation
\begin{align}  \label{gamma_matrices}
		\gamma^{0}=
		\begin{pmatrix}
			I & 0\\
			0 & -I
		\end{pmatrix},\ 
		\gamma^{i}
		=
		\begin{pmatrix}
			0 & \sigma^{i}\\
			-\sigma^{i} & 0
		\end{pmatrix}, \ 
		\gamma^{5}=
		\begin{pmatrix}
			0 & I\\
			I & 0
		\end{pmatrix},
\end{align}
with $\sigma^i$ the general Pauli matrices and $I$ the $2\times2$ unit matrix. Throughout this paper the metric signature is $(+,-,-,-)$.

In the preheating stage, when the expansion of the universe and the back-reaction of inflaton decays are neglected, 
the classical homogeneous solution for an inflaton of maximum amplitude $\phi_0$ follows from the free action $S_0[\phi]$:
\begin{align}
	\phi(t)=\phi_{0}\sin m_{\phi}t. \label{phi_t}
\end{align}
In the presence of the pseudoscalar interaction, the helicity of the fermions reverses each time when $\phi(t)$ crossing zero. 
As a result, particle production of a definite helicity is confined to brief intervals around the instants at which $\phi=0$. 
At these points, the interaction term $g\phi\bar\psi\mathrm{i}\gamma^5\psi$ vanishes so that helicity-asymmetric fermions generate, 
which allows the same effective-action formalism previously  developed for bosonic preheating \cite{effective_boson_preheating} to be applied.

The oscillating background $\phi(t)$ endows the Dirac field with a time-dependent frequency $\omega(t)$. 
The adiabaticity parameter $\dot\omega/\omega^2$ is strongly violated only when the inflaton passes through the minimum of its potential. 
These minima occur at the discrete times $t_j=j\pi/m_\phi$ ($j\in\mathbb{Z}$). 
Away from the minima one has $\dot\omega/\omega^2\ll1$, so that the occupation numbers of $\psi$ and $\bar\psi$ remain essentially constant. 
It is therefore sufficient to analyse particle creation in the immediate vicinity of each zero-crossing, where the inflaton may be linearised as
\begin{align}
	\phi(t)\approx(-1)^{j}\phi_{0}m_{\phi}(t-t_{j}),\quad t_{j}=\frac{j\pi}{m_{\phi}}. \label{phi_t_approx}
\end{align}
This local approximation greatly simplifies the evaluation of the effective action and yields analytic results.

In a flat spacetime, $\phi_0$ is a constant; while in an expanding universe the amplitude becomes time-dependent, namely $\phi_0(t)$. 
The linearisation \eqref{phi_t_approx} remains valid provided the fractional change of $\phi_0(t)$ over one oscillation period is small compared with the Hubble time $1/H$. 
In this section, these assumptions are applied throughout the preheating epoch, and in Sec. \ref{sec_extension}, we will also discuss the case of damping inflaton for the realistic physical purpose.

\subsection{One-loop effective action \label{subsec_oneloop}}

Since the non-perturbative decay of the inflaton produces Dirac particles $\psi$ and antiparticles $\bar\psi$ as final states, 
we evaluate the process by means of the one-loop effective action obtained after integrating out the fermionic fields:
\begin{align}
	\mathrm{e}^{\mathrm{i}S_{\mathrm{eff}}[\phi]}=\mathrm{e}^{\mathrm{i}S_{0}[\phi]}\int\mathcal{D}\bar{\psi}\,\mathcal{D}\psi\,
		\mathrm{e}^{\mathrm{i}S_{0}[\bar{\psi},{\psi}]+\mathrm{i}S_{Y}[\phi,\bar{\psi},\psi]},
\end{align}
with $\mathcal D$ the functional integration. 
Substituting the actions given in Eqs. \eqref{actions} yields  
\begin{align}
		\mathrm{e}^{\mathrm{i}S_{\mathrm{eff}}[\phi]}=\mathcal{N} &\mathrm{e}^{-\frac{\mathrm{i}}{2}\int\mathrm{d}^{4}x\,\phi(\partial_{\mu}\partial^{\mu}+m_{\phi}^{2})\phi}\cdot \nonumber\\
			&\qquad \det\left(\mathrm{i}\slashed{D}-m_{\psi}-\mathrm{i}g\phi\gamma^{5}+\mathrm{i}\varepsilon\right),  \label{exp_Seff}
\end{align}
where $\partial_\mu\partial^\mu=\partial_t^2-\nabla^2$, $\varepsilon>0$ is an infinitesimal convergence factor, $\slashed{D}=\gamma^\mu\partial_\mu$, $\mathcal{N}$ is a normalisation constant, 
and $\det(\hat{A})$ denotes the functional determinant of the operator $\hat{A}$. 
Taking the logarithm of both sides of Eq. \eqref{exp_Seff} gives
\begin{align}
	\mathrm{i}S_{\mathrm{eff}}[\phi]=&\ln\mathcal{N}-\frac{\mathrm{i}}{2}\int\mathrm{d}^{4}x\,\phi\left(\partial_{\mu}\partial^{\mu}+m_{\phi}^{2}\right)\phi(x) \nonumber\\
	&\qquad+\ln\det\left(\mathrm{i}\slashed{D}-m_{\psi}-\mathrm{i}g\phi\gamma^{5}+\mathrm{i}\varepsilon\right) \nonumber\\
    =&\mathrm{Tr}\left[\mathrm{tr}\ln\left(\mathrm{i}\slashed{D}-m_{\psi}-\mathrm{i}g\phi\gamma^{5}+\mathrm{i}\varepsilon\right)\right]+\mathrm{const.}    \nonumber\\
    =&\mathrm{Tr}\int\mathrm{d}^{4}x\,\langle x|\ln\left(\mathrm{i}\slashed{D}-m_{\psi}-\mathrm{i}g\phi\gamma^{5}+\mathrm{i}\varepsilon\right)|x\rangle \nonumber\\ &\qquad+\mathrm{const} .
\end{align}
In the second step we have used the identity $\ln\det\hat{A}=\operatorname{Tr}[\operatorname{tr}\ln\hat{A}]$. Here $\operatorname{Tr}$ stands for the trace over Dirac indices, 
while $\operatorname{tr}$ denotes the ordinary configuration-space integral $\int\mathrm{d}^4x\langle x|\cdots|x\rangle$.

Differentiating with respect to $m_\psi^2$ connects the expression to the Schwinger proper-time representation:
\begin{align} \label{Seff_dm2}
	&\frac{\partial\,\mathrm{i}S_{\mathrm{eff}}}{\partial m_{\psi}^{2}} \nonumber\\
	&=\frac{-1}{2m_{\psi}}\mathrm{Tr}\int\mathrm{d}^{4}x\,\left\langle x\left|\frac{1}{\mathrm{i}\slashed{D}-m_{\psi}-\mathrm{i}g\phi\gamma^{5}+\mathrm{i}\varepsilon}\right|x\right\rangle \nonumber\\
    &=\frac{-1}{2m_{\psi}}\operatorname{Tr}\int\mathrm{d}^{4}x\,\left\langle x\left|\frac{\mathrm{i}\slashed{D}-\mathrm{i}g\phi\gamma^{5}+m_{\psi}}
        {-\left(\slashed{D}-g\phi\gamma^{5}\right)^{2}-m_{\psi}^{2}+\mathrm{i}\varepsilon}\right|x\right\rangle \nonumber\\
    &=-\frac{1}{2}\operatorname{Tr}\int\mathrm{d}^{4}x\,\left\langle x\left|\frac{1}{-\left(\slashed{D}-g\phi\gamma^{5}\right)^{2}-m_{\psi}^{2}+\mathrm{i}\varepsilon}\right|x\right\rangle. 
\end{align}
In the last equality we have used the fact that the trace of an odd number of Dirac matrices vanishes. Introducing the Schwinger proper-time parameter $s$ then yields
\begin{align}   \label{Seff_s}
    &\mathrm{i}S_{\mathrm{eff}}= \nonumber\\
    &\frac{1}{2\mathrm{i}}\operatorname{Tr}\int_{0}^{\infty}\mathrm{d}s\int\mathrm{d}^{4}x\,\left\langle x\left|
        \mathrm{e}^{-\mathrm{i}\left[\left(\slashed{D}-g\phi\gamma^{5}\right)^{2}+m_{\psi}^{2}-\mathrm{i}\varepsilon\right]s}\right|x\right\rangle. 
\end{align}
Defining the following operator for convenience, 
\begin{align}
		\hat{H}=\left(\slashed{D}-g\phi\gamma^{5}\right)^{2}, \label{H}
\end{align}
and integrating with respect to $m_\psi^2$ finally produces the compact expression of the effective action
\begin{align}
	\mathrm{i}S_{\mathrm{eff}}=\frac{1}{2}\operatorname{Tr}\int_{0}^{\infty}\frac{\mathrm{d}s}{s}\,\mathrm{e}^{-\varepsilon s}
        \mathrm{e}^{-\mathrm{i}m_{\psi}^{2}s}\int\mathrm{d}^{4}x\,\langle x|\mathrm{e}^{-\mathrm{i}\hat{H}s}|x\rangle . \label{Seff_final}
\end{align}

\subsection{Eigenstates \label{subsec_eigen}}

Equation \eqref{Seff_final} shows that the effective action is determined by the amplitude $\langle x|\mathrm{e}^{-\mathrm{i}\hat{H}s}|x\rangle$ for a particle to return to the same configuration-space point after proper time $s$. 
Evaluating the configuration-space trace therefore requires the eigenvalues of the operator $\hat{H}$ defined in Eq. \eqref{H}, i.e., solutions of the eigenvalue problem $\hat{H}\Psi=E\Psi$.

Because sufficient fermion production occurs only near the bottoms of the inflaton potential, the eigenvalue equation is local to each production event: $\hat{H}\Psi_j=E^{(j)}\Psi_j$ for the $j$-th zero-crossing. 
Using the anti-commutation relations of gamma-matrices $\{\gamma^\mu,\gamma^\nu\}=2\eta^{\mu\nu}$ and $\{\gamma^\mu,\gamma^5\}=0$, 
and expanding the eigenfunction in momentum space as $\Psi_{j}=(2\pi)^{-3/2}\mathrm{e}^{\mathrm{i}\mathbf{k}\cdot\mathbf{x}}\varphi_{j}(t,k)$, one has the eigenequation 
\begin{align}
	&\left[\partial_{t}^{2}+g^{2}\phi_{0}^{2}m_{\phi}^{2}(t-t_{j})^{2}\right]\varphi_{j}= \nonumber\\
	&\qquad \left[E^{(j)}-k^{2}+(-1)^{j}\gamma^{0}\gamma^{5}g\phi_{0}m_{\phi}\right]\varphi_{j}.  \label{eiegn_phi_j}
\end{align}
This is the equation of an inverted harmonic oscillator. 
We introduce the orthonormal eigenvectors $\zeta_s^\lambda$ of the matrix $\gamma^0\gamma^5$, defined by the eigenvalue equation $\gamma^{0}\gamma^{5}\zeta^{\lambda}=\mathrm{i}\lambda\zeta^{\lambda}$, 
where $\lambda=\pm$ and $s$ labels the spin degrees of freedom. The eigenvalues are purely imaginary because $\gamma^0\gamma^5$ is antisymmetric. 
The system is degenerate with respect to spin; this degeneracy does not generate new physical effects and will therefore be suppressed in the subsequent discussion.
Then, decomposing $\varphi_j=f^{(j)}(t)\zeta^\lambda$ and introducing the dimensionless variables
\begin{align}
	&\omega_{0}^{2}=g\phi_{0}m_{\phi},\ \ \tau=\sqrt{\mathrm{i}}\,\omega_{0}(t-t_{j}),\nonumber\\ 
    &\kappa=\mathrm{i}\left[E/\omega_{0}^2-k^2/\omega_{0}^2+\mathrm{i}\lambda(-1)^{j}\right],  \label{parameters}
\end{align}
reduces Eq. \eqref{eiegn_phi_j} (omitting the label $j$) to the standard harmonic-oscillator equation
\begin{align}
	\frac{\mathrm{d}^{2}f}{\mathrm{d}\tau^{2}}+(\kappa-\tau^{2})f=0. \label{eiegn_f_j}
\end{align}
This is a standard harmonic oscillator equation. 
Imposing the natural boundary condition $f\to0$ as $\tau\to\infty$ quantises the eigenvalues to the odd integers $\kappa_n=2n+1$ ($n=0,1,2,\dots$). 
Consequently, the eigenvalues of the original operator are
\begin{align}
	E_{n,k,\lambda}^{(j)}=-\mathrm{i}\omega_{0}^{2}\left[2n+1-(-1)^{j}\lambda\right]\omega_{0}^2+k^{2}, \label{eigenvalue_E}
\end{align}
with $n=0,1,2,\cdots$, $\lambda=\pm$, and $k\geq0$. The corresponding eigenfunctions $f_n^{(j)}(t)$ satisfy the orthonormality relation
\begin{align}
		\int\mathrm{d}t f_{n}^{*}f_{n'}=\delta_{nn'},\ \text{with } n,n'=0,1,2,\cdots. \label{orthonormal_f}
\end{align}

The eigenvectors $\zeta^\lambda$ are not directly tied to physical observables; the eigenvectors of $\gamma^0$ themselves encode the helicity asymmetry. 
We therefore introduce the helical basis $\xi_{s_z}^h$ defined by $\gamma^{0}\xi_{s_z}^{h}=h\xi_{s_z}^{h}$, where $h=\pm$ labels the helical states and $s_z$ denotes the spin states. 
In the representation of the gamma matrices given in Eq.~\eqref{gamma_matrices}, these helical eigenvectors take the explicit form
\begin{align}
	\xi_{s_z}^{+}=(\chi_{s_z},\mathbf{0})^T,\qquad\xi_{s_z}^{-}=(\mathbf{0},\chi_{s_z})^T, \label{xi_h}
\end{align}
where $\chi_{s_z}$ are the rigenvectors of $\sigma_z$, with $\sigma_z\chi_{s_z}=s\chi_{s_z}$. 
They obey the orthonormality relations
\begin{align}
	(\xi_{s_z}^{h})^{\dagger}\xi_{s'_z}^{h'}=\delta_{hh'}\delta_{s_z s_z'},\quad\operatorname{Tr}\xi_{s_z}^{h}(\xi_{s'_z}^{h'})^{\dagger}=\delta_{hh'}\delta_{s_z s'_z}. \label{orthonormal_xi}
\end{align}
Using $\gamma^5\xi^\pm=\xi^\mp$, the helical vectors can be expressed as linear combinations of the $\zeta^\lambda$ basis: 
\begin{align}
		\xi_{s_z}^{\pm}=\frac{1}{\sqrt{2}}\zeta_{s_z}^{+}\mp\frac{\mathrm{i}}{\sqrt{2}}\zeta^-_{s_z}. \label{xi2zeta}
\end{align}
This transformation makes explicit the relation between the two bases and provides a linear way to translate between their respective representations.
A further identity, obtained by Taylor expansion, will be useful later: 
\begin{align}
	&\mathrm{e}^{\mathrm{i}\gamma^{0}\gamma^{5}z}\xi^{\pm}=\sum_{n=0}^{\infty}\frac{\left(\mathrm{i}\gamma^{0}\gamma^{5}z\right)^{n}}{n!}\xi^{\pm} \nonumber\\
    &=\sum_{n=0}^{\infty}\left(1+\mathrm{i}z\gamma^{0}\gamma^{5}+\frac{z^{2}}{2!}+\frac{z^{3}}{3!}\mathrm{i}\gamma^{0}\gamma^{5}+\cdots\right)\xi^{\pm} \nonumber\\
	&=\xi^{\pm}\cosh z\mp\mathrm{i}\xi^{\mp}\sinh z. \label{gamma5_zeta}
\end{align}
In the second step, we employ the relation $(\mathrm i\gamma^{0}\gamma^{5})^{2} = I$; and in the third step, we use $\gamma^{0}\gamma^{5}\xi^{\pm} = \mp\xi^{\mp}$. 
The exponential operator therefore mixes the two helical sectors through hyperbolic functions, providing a convenient equality for evaluating transition amplitudes between opposite-helicity states.

\subsection{Parametric resonance \label{subsec_para_resonance}}

Fermion production proceeds through a sequence of successive scattering events localised near the instants $t_j$. 
We now concentrate on a single production event (half an oscillation period), labelled by the index $j$. 
The completeness relation 
\begin{align}
	I=\sum_{n=0}^\infty \sum_{h=\pm} \int\mathrm{d}^3k |\mathbf{k},n,h\rangle\langle\mathbf{k},n,h|, \label{completeness}
\end{align}
is useful in the subsequent calculations. Here $\mathbf{k}$ denotes the momentum, $n$ the quantum number of the inverted harmonic oscillator in the time coordinate, and $h$ the helicity associated with the eigenvector $\xi^h$. 
Strictly speaking, the completeness sum should also run over the spin index $s_z$; however, this internal degree of freedom does not generate new physical effects and will therefore be omitted. 
Inserting the completeness relation \eqref{completeness} into effective action \eqref{Seff_final} yields, for the $j$-th production event,
\begin{widetext}
\begin{align}
	\mathrm{i}S_{\mathrm{eff}}^{(j)}&=\frac{1}{2}\operatorname{Tr}\sum_{n,n'=0}^{\infty}\sum_{h,h'=\pm}\int \mathrm{d}^{3}k\int \mathrm{d}^{3}k'\int_{0}^{\infty}\frac{\mathrm{d}s}{s}\mathrm{e}^{-\varepsilon s}\mathrm{e}^{-\mathrm{i}m_{\psi}^{2}s}
	\int \mathrm{d}^{4}x\,\langle x|\mathbf{k},n,h\rangle\langle\mathbf{k},n,h|e^{-\mathrm{i}\hat{H}s}|\mathbf{k}',n',h'\rangle\langle\mathbf{k}',n',h'|x\rangle.  \label{Seff_completeness}
\end{align}
According to the relations $\langle\mathbf{k}|\mathbf{x}\rangle=(2\pi)^{-3/2}\mathrm{e}^{\mathrm{i}\mathbf{k}\cdot\mathbf{x}}$ and $\langle\mathbf{k},n,h|\mathbf{k}',n',h'\rangle=\delta^3(\mathbf{k}-\mathbf{k}')\delta_{nn'}\delta_{hh'}$, 
together with the eigenvalue \eqref{eigenvalue_E}, the orthonormal relations \eqref{orthonormal_f} and \eqref{orthonormal_xi}, and linear transform \eqref{xi2zeta}, we then have
\begin{align}
	\mathrm{i}S_{\mathrm{eff}}^{(j)} = &\frac{1}{2}\int\frac{\mathrm{d}^{3}k}{(2\pi)^{3}}\int\mathrm{d}^{3}x\int_{0}^{\infty}\frac{\mathrm{d}s}{s}\sum_{n=0}^{\infty}\sum_{\lambda=\pm} 
        \mathrm{e}^{-\varepsilon s}\mathrm{e}^{-\mathrm{i}(k^{2}+m_{\psi}^{2})s}\mathrm{e}^{-\omega_{0}^{2}[2n+1-(-1)^{j}\lambda]s} \nonumber\\
    =& \frac{1}{2}\int\frac{\mathrm{d}^{3}k}{(2\pi)^{3}}\int\mathrm{d}^{3}x\int_{0}^{\infty}\frac{\mathrm{d}s}{s}\mathrm{e}^{-\varepsilon s}\mathrm{e}^{-\mathrm{i}(k^{2}+m_{\psi}^{2})s}\frac{\cosh\omega_0^2s} {\sinh\omega_0^2s}. 
\end{align}
\end{widetext}
In the continuum limit of statistical mechanics, it is valid for the replacement
\begin{align}
	\frac{1}{(2\pi)^3}\int\mathrm{d}^3k\int\mathrm{d}^3x \longleftrightarrow \sum_{\mathbf{K}},
\end{align}
so that
\begin{align}
	\mathrm{i}S_{\mathrm{eff}}^{(j)} =\frac{1}{2}\sum_{\mathbf{k}}\int_{0}^{\infty}\frac{\mathrm{d}s}{s}\mathrm{e}^{-\varepsilon s}\mathrm{e}^{-\mathrm{i}(k^{2}+m_{\psi}^{2})s}\frac{\cosh\omega_{0}^{2}s}{\sinh\omega_{0}^{2}s}. 
\end{align}
The effective action is therefore an incoherent sum over momentum modes. It is convenient to introduce the mode-resolved effective action $S_j(\mathbf{k})$ through 
\begin{align}
	S_{\mathrm{eff}}^{(j)}=\sum_{\mathbf{k}}S_j(\mathbf{k}).
\end{align}
We now evaluate $S_j(\mathbf{k})$ by performing the remaining Schwinger proper-time integral for a given momentum $\mathbf{k}$.

The Schwinger proper-time integral is evaluated with the residue theorem. The integrand diverges at $s=0$, and the singularity at this point is not a simple pole. 
We remove the divergence in a Lorentz-invariant manner by subtracting the $s=0$ pole. 
This subtraction is equivalent to the introduction of a counter-term that depends on the coupling $g$ and the amplitude $\phi_0$ but is independent of the external momentum \cite{Schwinger_time1,Schwinger_time2}. 
After regularisation the integral becomes
\begin{align}
	\mathrm{i}S_j(\mathbf{k})=&\frac{1}{2}\int_{0}^{\infty}\frac{\mathrm{d}s}{s}\mathrm{e}^{-\varepsilon s}\mathrm{e}^{-\mathrm{i}(k^{2}+m_{\psi}^{2})s}
	   \left(\frac{\cosh\omega_{0}^{2}s}{\sinh\omega_{0}^{2}s}-\frac{1}{\omega_{0}^{2}s}\right).  \label{Seff_regularization}
\end{align}
Changing variable to $z=\omega_0^2 s$ and rotating the contour into the Euclidean domain ($z\to-\mathrm{i}z$) yields 
\begin{align}
	\mathrm{i}S_j(\mathbf{k})=\frac{1}{2}\int_{0}^{\infty}\frac{\mathrm{d}z}{z}\mathrm{e}^{\mathrm{i}\varepsilon z}\mathrm{e}^{-(k^{2}+m_{\psi}^{2})z/\omega_{0}^{2}}\left(\cot z-\frac{1}{z}\right). 
\end{align}
The integrand is analytic in the complex $z$-plane except for simple poles at $z_n=n\pi$ ($n=1,2,3,\dots$) on the positive real axis, which coincide with the zeros of $\sin z$ (the apparent singularity at $s=0$ is removable). 
The residues of $\cot z$ at these poles are unity, i.e., $\operatorname{Res}(\cot z,z_n=n\pi)=1$. 
The infinitesimal $\mathrm{i}\varepsilon$ prescription shifts the poles slightly into the lower half-plane. 
Consequently, the integral along the positive real axis is equal to its principal-value contribution minus $\mathrm{i}\pi$ times the sum of the residues:
\begin{align}
	\mathrm{i}S_j(\mathbf{k})&=-\frac{\mathrm{i}\pi}{2}\sum_{n=1}^{\infty}\frac{1}{n\pi}\exp\left(-\frac{n\pi(k^{2}+m_{\psi}^{2})}{\omega_{0}^{2}}\right) \nonumber\\
	&=\frac{\mathrm{i}}{2}\ln\left[1-\exp\left(-\frac{\pi(k^{2}+m_{\psi}^{2})}{g\phi_{0}m_{\phi}}\right)\right].   \label{S_j}
\end{align}

The vacuum-to-vacuum transition amplitude, also named the vacuum persistence probability, gives the relative probability that no fermion pair is produced during the half-oscillation interval of the $j$-th production event:
\begin{align}
	&|\langle\mathrm{out}|\mathrm{in}\rangle|^2 \nonumber\\
    &=\prod_{\mathbf{k}}|P_{\mathrm{vac}}^{(j)}(\mathbf{k})|^{2}=\left|\mathrm{e}^{\,\mathrm{i}S_{\mathrm{eff}}^{(j)}}\right|^{2}=\mathrm{e}^{-2\operatorname{Im}S_{\mathrm{eff}}^{(j)}}\notag\\
		&=\mathrm{e}^{-2\sum_{\mathbf{k}}\operatorname{Im}S_j(\mathbf{k})}=\prod_{\mathbf{k}}\mathrm{e}^{-2\operatorname{Im}S_j(\mathbf{k})}. 
\end{align}
In writing the total persistence probability as a product over modes, we have assumed that distinct momentum modes contribute independently. 
The vacuum persistence probability for a single mode $\mathbf{k}$ is therefore
\begin{align}
    |P_{\mathrm{vac}}^{(j)}(\mathbf{k})|^2=\mathrm{e}^{-2\operatorname{Im}S_j(\mathbf{k})}. \label{P_j_k}
\end{align}
According to Eq. \eqref{S_j}, the imaginary part of the effective action is therefore
\begin{align}
		&\operatorname{Im}S_j(\mathbf{k}) \nonumber\\
		&=-\operatorname{Re}\frac{1}{2}\int_{0}^{\infty}\frac{\mathrm{d}z}{z}\mathrm{e}^{\mathrm{i}\varepsilon z}\mathrm{e}^{-(k^{2}+m_{\psi}^{2})z/\omega_{0}^{2}}\left(\cot z-z^{-1}\right) \nonumber\\
		&=-\frac{1}{2}\ln\left[1-\exp\left(-\frac{\pi(k^{2}+m_{\psi}^{2})}{g\phi_{0}m_{\phi}}\right)\right].  \label{Seff_Im_final}
\end{align}
Inserting this result into Eq. \eqref{P_j_k}, the vacuum persistence probability for $k$-mode is 
\begin{align}
    |P_{\mathrm{vac}}^{(j)}(\mathbf{k})|^2=1-\exp\left( -\frac{\pi(k^{2}+m_{\psi}^{2})}{g\phi_{0}m_{\phi}} \right). \label{P_j_k_final}
\end{align}
Thus the probability of creating fermion pairs reads
\begin{align}
	\Gamma_{\mathbf{k}}(\phi\rightarrow\bar{\psi}\psi)=1-|P_{\mathrm{vac}}(\mathbf{k})|^{2}=\mathrm{e}^{-\frac{\pi(k^{2}+m_{\psi}^{2})}{g\phi_{0}m_{\phi}}}.  \label{p_creat_pseudo}
\end{align}
Equation \eqref{p_creat_pseudo} reproduces the Gaussian spectrum obtained previously for bosonic preheating \cite{effective_boson_preheating}, scalar Yukawa preheating \cite{preheating_fermion1} and pseudoscalar Yukawa preheating \cite{Yukawa_preheating}.

Because Pauli blocking forbids multipair production in a given fermionic mode, the tunneling probability for fermions receives no contribution from multi-instanton or anti-instanton configurations, which is unlike the bosonic case. 
According to the instanton interpretation \cite{instanton}, the vacuum-to-vacuum transition is controlled by the imaginary part of the effective action. 
The fermion-pair production rate per unit phase-space volume after a single production event is therefore twice the imaginary part of the effective action: 
\begin{align}
	w&=-\frac{1}{V}\sum_{\mathbf{k}}\ln\left(1-\mathrm{e}^{-\frac{\pi (k^2+m^2_\psi)}{\omega_{0}^2}}\right) \nonumber\\
	&=\frac{1}{2\pi^2}\int\mathrm{d}kk^2\sum_{n=1}^{\infty}(-1)^{n-1}\mathrm{e}^{-{n\pi(k^2+m^2_\psi)}/{\omega_{0}^2}} \nonumber\\
	&=\frac{\omega_{0}^3}{(2\pi)^3}\sum_{n=1}^{\infty}\frac{(-1)^{n-1}}{n^{3/2}}\exp\left(-\frac{n\pi m^2_\psi}{\omega_{0}^2}\right) \nonumber\\
	&=-\frac{\omega_{0}^3}{(2\pi)^3}\mathrm{Li}_{\frac{3}{2}}\left(-\exp\left(-\frac{\pi m^2_\psi}{g\phi_0 m_\phi}\right)\right),
\end{align}
where $\mathrm{Li}_s(z)$ is the polylogarithm function. 

The analysis presented in this section is restricted to a single production event and does not yet address the succession of scatterings that occur as the inflaton continues to oscillate about the minimum of its potential. 
Moreover, the effective action $S_{\mathrm{eff}}^{(j)}$ constructed above is the total phase accumulated after one half-period and does not resolve the individual helical contributions. 
Because the pseudoscalar Yukawa interaction mixes the two helical sectors, seen as Eq. \eqref{gamma5_zeta}, 
a separate treatment of the helicity-dependent amplitudes is required in order to describe the successive generation of helicity-asymmetric fermions during the preheating epoch. 

\section{Helical asymmetry \label{sec_heli_asymmetry}}

The vacuum-to-vacuum amplitude after a single production event (half an oscillation) can be rewritten by projecting onto the helical basis $\{\xi_s^h\}$ introduced in Eqs. \eqref{xi_h}–\eqref{xi2zeta}. 
Starting from Eq. \eqref{Seff_completeness} and summing over helical states, the effective action takes the form
\begin{align}
    &\mathrm{e}^{\mathrm{i}S_j(\mathbf{k})}  \nonumber\\
	&=\exp\left[\frac{1}{4}\mathrm{Tr}\int_{0}^{\infty}\frac{\mathrm{d}s}{s}\mathrm{e}^{-\varepsilon s}\mathrm{e}^{-\mathrm{i}(k^{2}+m_{\psi}^{2})s} \frac{1}{{\sinh\omega_0^2s}}\right. \nonumber\\
	&\qquad\quad \left.\sum_{h,h'=\pm}|h\rangle\langle h|\mathrm{e}^{\mathrm{i}(-1)^{j}\omega_{0}^{2}s\gamma^0\gamma^5}|h'\rangle\langle h'|\right].
\end{align}
Notice the identity $\exp\operatorname{Tr}\mathbf{T}=\det\exp\mathbf{T}$, and we then have
\begin{align}
    \mathrm{e}^{\mathrm{i}S_j(\mathbf{k})}=\det\exp\mathbf{T}^{(j)},  \label{Seff_det_exp}
\end{align}
where $\mathbf{T}^{(j)}$ is a $2\times2$ matrix by tracing over the spin states. Here, $\mathbf{T}^{(j)}$ is named transfer matrix for $j$-th production event. Its diagonal elements are
\begin{subequations} 
\begin{align}
	&T_{11}^{(j)}(\mathbf{k})=T_{22}^{(j)}(\mathbf{k}) \nonumber\\
	&=\frac{1}{4}\int_{0}^{\infty}\frac{\mathrm{d}s}{s}\frac{\mathrm{e}^{-\varepsilon s}\mathrm{e}^{-\mathrm{i}(k^{2}+m_{\psi}^{2})s}}{\sinh\omega_{0}^{2}s}
        \langle h=+|\mathrm{e}^{\mathrm{i}(-1)^{j}\omega_{0}^{2}s\gamma^{0}\gamma^{5}}|+\rangle \nonumber\\
	&=\frac{1}{4}\int_{0}^{\infty}\frac{\mathrm{d}s}{s}\mathrm{e}^{-\varepsilon s}\mathrm{e}^{-\mathrm{i}(k^{2}+m_{\psi}^{2})s}\frac{\cosh\omega_{0}^{2}s}{\sinh\omega_{0}^{2}s}, \label{T1122}
\end{align} 
and the off-diagonal elements read
\begin{align}
	&T_{12}^{(j)}(\mathbf{k})=-T_{21}^{(j)}(\mathbf{k}) \nonumber\\
	&=\frac{1}{4}\int_{0}^{\infty}\frac{\mathrm{d}s}{s}\frac{\mathrm{e}^{-\varepsilon s}\mathrm{e}^{-\mathrm{i}(k^{2}+m_{\psi}^{2})s}}{\sinh\omega_{0}^{2}s}
	   \langle+|\mathrm{e}^{\mathrm{i}(-1)^{j}\omega_{0}^{2}s\gamma^{0}\gamma^{5}}|-\rangle  \nonumber\\
	&=\frac{\mathrm{i}}{4}(-1)^{j}\int_{0}^{\infty}\frac{\mathrm{d}s}{s}\mathrm{e}^{-\varepsilon s}\mathrm{e}^{-\mathrm{i}(k^{2}+m_{\psi}^{2})s}.   \label{T1221}
\end{align}
In the last step of the above equations, the helicity-mixing identity \eqref{gamma5_zeta} has been applied. 
\end{subequations}

This form reveals two important features.
The equal diagonal elements indicate that the two helical sectors have identical diagonal contributions in each production event.
While the alternating off-diagonal elements show the mixing between the two sectors reverses sign from one production event to the next, since the direction of time propagation is reversed. 
These off-diagonal terms are therefore directly responsible for the generation of a net helical asymmetry. 

After $N$ successive production events, the accumulated vacuum-to-vacuum amplitude in a given momentum mode is
\begin{align}
	P_N(\mathbf{k})&=\exp\left( \sum_{j=1}^N \mathrm{i}S_j(\mathbf{k})\right)=\prod\limits_{j=1}^{N}\mathrm{e}^{\mathrm{i}S_j(\mathbf{k})}    \nonumber\\
    &=\prod\limits_{j=1}^{N}\det\exp\mathbf{T}^{(j)}(\mathbf{k})=\det\prod\limits_{j=1}^{N}\exp\mathbf{T}^{(j)}(\mathbf{k}),    \label{PN_prodect}
\end{align}
where we used relation \eqref{Seff_det_exp} in the third step. 
Because the transfer matrices commute, $[\mathbf{T}^{(j+1)},\mathbf{T}^{(j)}]=0$, the product can be rewritten as a single exponential of the summed generators: 
\begin{align}
    P_N(\mathbf{k})=\det\exp\sum\limits_{j=1}^{N}\mathbf{T}^{(j)}(\mathbf{k}). \label{PN_exp_sum}
\end{align}
Thus the total amplitude after $N$ scatterings is completely determined by the coherent accumulation of the individual transfer matrices.

We now examine the generation of helical asymmetry. For any two successive production events, one finds
\begin{align}
	\mathbf{T}^{(j)}+\mathbf{T}^{(j+1)}=\begin{pmatrix}2T_{11}^{(j)} & 0\\0 & 2T_{22}^{(j)}\end{pmatrix}. 
\end{align}
Thus, after an even number of events the cumulative transfer matrix is strictly diagonal and the net mixing between the two helical sectors vanishes. 
Moreover, the equality $T_{11}^{(j)}=T_{22}^{(j)}$ implies that the two helicities remain equally populated. 
This equality is a direct consequence of the temporal symmetry of the undamped oscillation $\phi(t)=\phi_0\sin(m_\phi t)$ about each zero-crossing $t_j$. 
As a result, when the number of production events is even the symmetry of the background guarantees that the two helical sectors are produced with identical probabilities.

When the number of events is odd, $N=2n+1$, the residual off-diagonal parts survive 
\begin{align} \label{T_odd}
	\sum_{j=1}^{2n+1}\mathbf{T}^{(j)}=\begin{pmatrix}
			(2n+1)T_{11}^{(1)} & T_{12}^{(1)}\\
			-T_{12}^{(1)} & (2n+1)T_{22}^{(1)}
	\end{pmatrix}.
\end{align}
The incomplete cancellation of the mixing terms arises the possibility of a net helical asymmetry. 
The final helicity composition of the fermion population therefore depends sensitively on the parity of the number of production events as well as on the detailed structure of the transfer matrix.
In particular, an odd number of production events can give rise to a helical asymmetry if the off-diagonal contributions lead to unequal populations of the two helical sectors. 
The resulting asymmetry therefore depends not only on the parity of the number of production events but also on the detailed structure of the transfer matrix and its action on the initial fermionic state.

To quantify the asymmetry, we diagonalise the accumulated generator. 
Assume $\lambda_\pm^{(N)}$ are the eigenvalues of $\sum_{j=1}^{N}\mathbf{T}^{(j)}$.  
For odd $N=2n+1$, therefore, 
\begin{align}
	\lambda_{\pm}^{(2n+1)}=(2n+1)T_{11}^{(1)}(\mathbf{k})\pm \mathrm{i}T_{12}^{(1)}(\mathbf{k}). 
\end{align}
The total vacuum-to-vacuum amplitude may then be written
\begin{align}
	P_{N}(\mathbf{k})=\exp\left(\lambda_{+}^{(N)}+\lambda_{-}^{(N)}\right). 
\end{align}
The quantities $\lambda_{\pm}^{(N)}$ are precisely the accumulated phases associated with each helical sector. 
Decomposing the effective action into helical components, $\mathrm{i}S_j(\mathbf{k})=\mathrm{i}S_j^{(+)}+\mathrm{i}S_j^{(-)}$, with $\sum_{j=1}^N S_j^{(h)}=-\mathrm{i}\lambda_h^{(N)}$, 
one obtains the imaginary parts of helical effective action after odd numbers of production:
\begin{align}
	&-2\operatorname{Im}S_{2n+1}^{(h=\pm)}(\mathbf{k})=2\operatorname{Re}\lambda_{h}^{(2n+1)} \nonumber\\
	&=\frac{1}{2}\operatorname{Re}\int_{0}^{\infty}\frac{\mathrm{d}z}{z}\mathrm{e}^{\mathrm{i}\varepsilon z}\mathrm{e}^{-(k^{2}+m_{\psi}^{2})z/\omega_{0}^{2}}\left(\cot z-z^{-1}\right) \nonumber\\
    &\quad \pm\frac{1}{2}\operatorname{Re}\int_{0}^{\infty}\frac{\mathrm{d}z}{z}\mathrm{e}^{-\varepsilon z}\left[\mathrm{e}^{-\mathrm{i}(k^{2}+m_{\psi}^{2})z/\omega_0^2}-\mathrm{e}^{-\mathrm{i} \mu z}\right],
\end{align}
where $\mu$ is the renormalization scale.
The diagonal contribution has already been evaluated in Eq. \eqref{Seff_Im_final}; the off-diagonal integral is regularised by the same $\mathrm{i}\varepsilon$ prescription (or equivalently by subtraction of a reference mass term). 

Because Pauli blocking prevents multi-pair production in a given fermionic mode, the total tunneling probability for each helical mode is simply
\begin{align}
	\Gamma^{(\pm)}(k)=1-|P_N^{(\pm)}(\mathbf{k})|^2=1-\mathrm{e}^{2\operatorname{Re}\lambda_{\pm}^{(N)}}. \label{Gamma_pm_k}
\end{align}
Equivalently, within the instanton interpretation \cite{instanton}, the helicity-resolved production rate per unit phase space after $N$ successive events is twice the imaginary part of the accumulated effective action:
\begin{align}
	w_h^{(N)}=2\operatorname{Im}\sum_{j=1}^N S_j^{(h)}(\mathbf{k})=-2\operatorname{Re}\lambda_{h}^{(N)}.
\end{align}

The conclusions of the present section rely on the undamped oscillatory background of Eq. \eqref{phi_t}, which is symmetric under reflection about each zero-crossing. 
This symmetry produces a coherent cancellation of the helical mixing after every even number of events. 
In order to obtain a sustained net helical asymmetry it is therefore necessary to introduce a damped oscillatory background that breaks the temporal symmetry about the potential minima.

\section{Extension of the model \label{sec_extension}}

\subsection{Scalar Yukawa interaction \label{subsec_scalarYukawa}}

In Sec. \ref{sec_preheating} and \ref{sec_heli_asymmetry}, the results are based the pseudoscalar Yukawa action \eqref{S_Y}, showing a helical asymmetry after event number of scattering events. 
We now consider an alternative model with scalar Yukawa coupling between the scalar inflaton and Dirac fields, whose action is given by
\begin{align}
    S_{Y}[\phi,\bar{\psi},\psi]=-\int\mathrm{d}^{4}x\,g\phi\,\bar{\psi}\psi. \label{S_scalar_Y}
\end{align}
The oscillating mode of inflaton is still considered as Eq. \eqref{phi_t}. 
Repeating the calculations in Sec. \ref{subsec_oneloop}, the effective action with respect to the functional integral of $\bar\psi$ and $\psi$ reads
\begin{align}
	&\frac{\partial \mathrm{i}S_{\mathrm{eff}}}{\partial m_{\psi}^{2}} \nonumber\\
	&=-\frac{1}{2m_{\psi}}\operatorname{Tr}\int \mathrm{d}^{4}x\,\left\langle x\left|\frac{1}{\mathrm{i}\slashed{D}-g\phi-m_{\psi}+\mathrm{i}\varepsilon}\right|x\right\rangle \nonumber\\
	&=-\frac{1}{2}\operatorname{Tr}\int \mathrm{d}^{4}x\,\left\langle x\left|\frac{1}{-(\slashed{D}+\mathrm{i}g\phi)^{2}-m_{\psi}^{2}+\mathrm{i}\varepsilon}\right|x\right\rangle.
\end{align}
With integrating over $m_\psi^2$, effective action is still expressed as Eq. \eqref{Seff_s}, but the operator becomes
\begin{align}
	\hat{H}=(\slashed{D}+\mathrm{i}g\phi)^{2}. 
\end{align}
The precise form of $\hat H$ differs from the pseudoscalar case by the absence of the $\gamma^5$ matrix, but the subsequent spectral analysis proceeds analogously.
Near each zero-crossing $t_j=j\pi/m_\phi$ one linearizes $\phi(t)\approx(-1)^j\phi_0 m_\phi(t-t_j)$. 
After Fourier transformation and projection onto the helical basis $\{\xi_h^s\}$ defined in Eqs. \eqref{xi_h} - \eqref{xi2zeta}, the eigenvalue problem for $\hat H$ decouples into two independent sectors labelled by helicity $h=\pm$. 
Then, the analytical expression of the effective action with regularization is exactly the same as Eq. \eqref{Seff_regularization}. 
As well, the fermion pair-creation probability $\Gamma_\mathbf{k}$ is also coincide with the pseudoscalar case, shown as expression \eqref{p_creat_pseudo}, since the eigenvalues of $\hat H$ are the same in both cases. 

However, the critical difference between the scalar and pseudoscalar preheating is the transfer matrix $\mathbf{T}^{{j}}$. 
In the scalar case, the transfer matrix across the $j$-th production event is strictly diagonal:
\begin{align} 
	\mathbf{T}^{(j)}=\begin{pmatrix}
			T_{11}^{(1)} & 0\\
			0 & T_{22}^{(1)}
	\end{pmatrix}.
\end{align}
for any $j=1,2,3,\cdots$. The expressions of $T_{11}^{(1)}$ and $T_{22}^{(1)}$ are both the exactly the same as the on in Eq. \eqref{T1122}, and $T_{12}^{(1)}=T_{21}^{(1)}=0$.
The off-diagonal elements vanish identically because the scalar interaction does not mix the two helical eigenvectors of $\gamma^0$. 
After an arbitrary number $N$ of successive production events the accumulated amplitude therefore remains diagonal in the helical basis
\begin{align}
    P_N(k)&=\det\exp\left(\sum_{j=1}^N\mathbf{T}^{(j)}(k)\right)\nonumber\\ &=\exp\left(N\,T_{11}^{(1)}(k)\right)\exp\left(N\,T_{22}^{(1)}(k)\right).
\end{align}
Since $T_{11}^{(1)}=T_{22}^{(1)}$, the two helicities acquire identical phases. 
The corresponding pair-production probabilities are therefore equal, and no net helical asymmetry is generated, irrespective of the parity of $N$.
The crucial distinction lies solely in the absence of helical mixing: the scalar Yukawa interaction preserves the equality of the two helicity populations at every stage of the preheating process. 
Thus the generation of a net fermionic helicity asymmetry is a distinctive signature of the pseudoscalar coupling and does not occur for a pure scalar Yukawa interaction.

\subsection{Decay oscillation of inflaton \label{subsec_decay_oscillation}}

In Eq. \eqref{phi_t}, we introduce an oscillation mode without damping and exhabits a symmetry about time $t$ around $t_j$. 
Because this solution is symmetric under $t-t_j\to t_j-t$ about every zero-crossing $t_j$, the off-diagonal (helicity-mixing) elements of the transfer matrices cancel exactly after an even number of productions, 
and the residual mixing after an odd number of productions is subsequently removed by the next even event. Consequently a sustained net helical asymmetry cannot accumulate.

A realistic post-inflationary evolution, however, includes the Hubble damping of the inflaton amplitude. After the first few oscillations the homogeneous background is well approximated by
\begin{align}
    \phi(t)=\phi_{0}\frac{\pi\sin m_{\phi}t}{m_{\phi}t}. \label{phi_t_damped}
\end{align}
The amplitude now decreases monotonically, so the linearised damping at successive zero-crossings. 
In this situation, the non-adiabatic region approximates
\begin{align}
	\phi(t)\approx(-1)^{j}\phi_{0}m_{\phi}(t-t_{j})/j,\quad t_{j}={j\pi}/{m_{\phi}}. \label{phi_t_damped_approx}
\end{align}
The time-reflection symmetry of $\phi(t)$ about each zero-crossing $t_j$ is now broken. As a result, helical symmetry may be also violated after any number of oscillations.

Following the effective-action procedure developed in Sec. \ref{sec_heli_asymmetry}, we denote the matrix associated with the $j$-th production event by $\mathbf T^{(j)}$, defined in Eq. \eqref{Seff_det_exp}. 
Its diagonal components have the form
\begin{subequations} \label{AAA}
\begin{align}
	T_{11}^{(j)}(\mathbf{k})&=T_{22}^{(j)}(\mathbf{k}) \nonumber\\
	&=\frac{1}{4}\int_{0}^{\infty}\frac{\mathrm{d}z}{z}\mathrm{e}^{-\varepsilon z}\mathrm{e}^{-\mathrm{i}(k^{2}+m_{\psi}^{2})z/\omega_{j}^{2}}\frac{\cosh z}{\sinh z}, \label{T1122_damped}
\end{align} 
with $\omega_{j}^{2}=g\phi_0 m_\phi/j$, the same regularization prescription as in Sec. \ref{sec_preheating} is applied;
and the off-diagonal terms are expressed as
\begin{align}
	T_{12}^{(j)}(\mathbf{k})&=-T_{21}^{(j)}(\mathbf{k}) \nonumber\\
	&=\frac{\mathrm{i}}{4}(-1)^{j}\int_{0}^{\infty}\frac{\mathrm{d}z}{z}\mathrm{e}^{-\varepsilon z}\mathrm{e}^{-\mathrm{i}(k^{2}+m_{\psi}^{2})z/\omega_{j}^{2}}.   
\end{align} 
\end{subequations}
The off-diagonal components are obtained from the corresponding helicity-transition matrix elements, with the alternating sign associated with the crossing direction. 
However, the $j$-dependent parameter $\omega_{j}^{2}$ means the non-commutation relation $[\mathbf{T}^{(j+1)},\mathbf{T}^{(j)}]\neq0$. 
It implies that the cumulated amplitude after $N$-successive scattering can not be written in the form of the summation of each transfer matrices, as shown in Eq. \eqref{PN_exp_sum};
instead, its expression should be in the form of Eq. \eqref{PN_prodect}. 

Define the transfer matrix $\mathbf{R}^{(N)}$ as a function of momentum $k$, with expression
\begin{align}
    \exp\mathbf{R}^{(N)}\equiv\prod\limits_{j=1}^{N}\exp\mathbf{T}^{(j)}(\mathbf{k}), \label{expR}
\end{align}
which means the transfer magnitude after $N$-times successive scattering events. 
Note that matrix $\mathbf{R}^{(N)}$ is off-diagonal for any $N\geq1$, indicating that the fermion pair-production rates for different helicity are always distinguishable. 
Thus, based on the discussions in this paper, fermionic preheating with helical asymmetry need two necessary conditions, one is the interaction that couples the different helical fermions, such as the pseudoscalar Yukawa interaction, 
the other is the dynamical asymmetry that breaks the time symmetry, such as the damped oscillation of inflaton. 

\subsection{Bogoliubov coefficients and production probabilities \label{subsec_Bog}}

\begin{figure*}
	\centering
	\includegraphics[width=\linewidth]{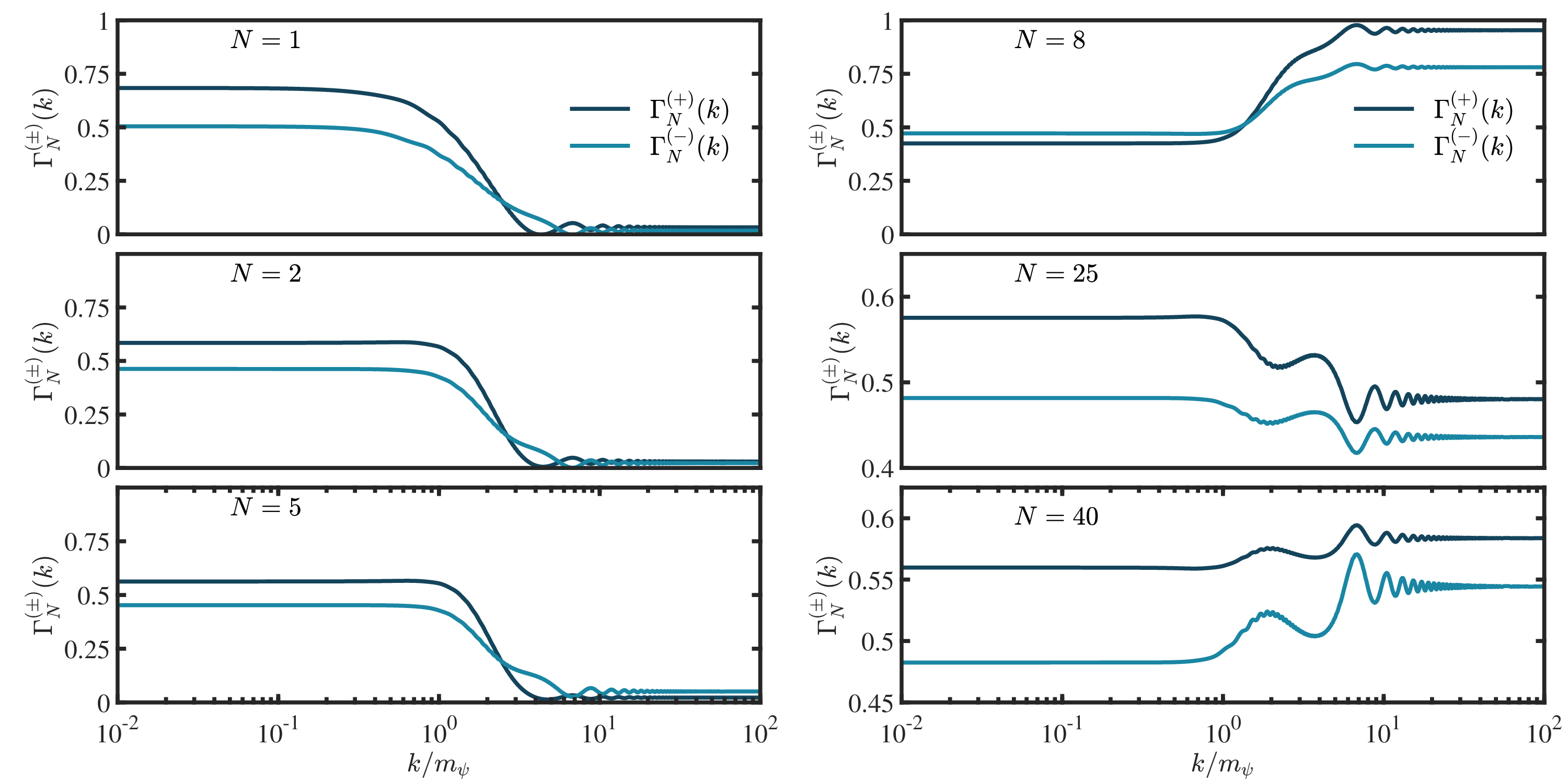}
	\caption{Helicity-dependent production probabilities $\Gamma_N^{(\pm)}(k)$ versus dimensionless momentum $k/m_\psi$ for successive production events with $N=1,3,5,8,25$ and $40$. The parameter is set $\omega_0^2=5\pi m_\psi^2$.  }
	\label{Gamma_Fig}
\end{figure*}

The damped background breaks the exact temporal symmetry of the undamped oscillation and changes the nonadiabatic evolution at successive zero crossings. 
It can therefore modify the relative phases and magnitudes of the accumulated production amplitudes. 
Whether this leads to a nonvanishing $\Delta n_{\mathbf k}$, however, must be established by calculating the outgoing states and comparing the two helicity populations. 
The damping of the inflaton alone is not sufficient to establish helicity asymmetry; the interaction structure, the evolution of the helicity channels, and the initial state must also be taken into account. 

Fermionic preheating previously are investigated via WKB approximation, which are in terms of the Bogoliubov coefficients. 
In expanding universe, with the application of the "random walk recipe", the Bogoliubov coefficients between two successive productions are given by
\begin{align}  \label{Bog_coeffi}
    \begin{pmatrix}
    |\alpha_n^{(r)}|^2 \\[1mm]
    |\beta_n^{(\bar r)}|^2
    \end{pmatrix}
    =
    \begin{pmatrix}
    |F_n^{(r)}|^2 & |H_n^{(r)}|^2 \\[1mm]
    |H_n^{(r)}|^2&  |F_n^{(r)}|^2
    \end{pmatrix}
    \begin{pmatrix}
    |\alpha_{n-1}^{(r)}|^2 \\[1mm]
    |\beta_{n-1}^{(\bar r)}|^2
    \end{pmatrix}.
\end{align}
Here, $\alpha_n^{(r)}$ and $\beta_n^{(r)}$ are the values of the Bogolyubov coefficients after the $n$-th production, 
and $F_n^{(r)}$ and $H_n^{(r)}$ are transfer amplitudes between two productions. 
In this equation, a fermion particle may come from two ways: either as article from
the in-vacuum state with amplitude $F_n$ or through a transition from an antiparticle into a particle with amplitude $H_n$. 
The same interpretation applies to the creation of an anti-fermion particle. 
For the fermionic preheating with scalar Yukawa interaction \cite{preheating_fermion1}, as introduced in Sec. \ref{subsec_scalarYukawa}, all the parameters in Eq. \eqref{Bog_coeffi} are independent on the helical states $r$, 
and follows the relations $|F_n|^2=1-\mathrm e ^{-\pi p_n^2}$, $|H_n|^2=\mathrm e ^{-\pi p_n^2}$ and $|F_n|^2+|H_n|^2=1$, with $p_n^2\propto k^2$. 
This model generates the equal helicities after series of productions. 
For the model with pseudoscalar Yukawa interaction \cite{Yukawa_preheating}, as introduced in Sec. \ref{subsec_decay_oscillation}, the Bogolyubov coefficients $\alpha_n^{(r)}$ and $\beta_n^{(r)}$ are helical dependent, 
and helical asymmetry spontaneously arisen after series of successive production events. 
In this model, the parameters satisfy $|F_n^{(r)}|^2 = 1 - \mathrm{e}^{-\pi \kappa_{r,n}^2}$ and $|H_n^{(r)}|^2 =\mathrm{e}^{-\pi \kappa_{r,n}^2}$, where $\kappa_{r,n}^2$ is helical dependent. 
They both have a similar form as shown in Eq. \eqref{P_j_k_final}. The slight differences may come from their approximate methods used in the calculations. 

In the analysis above, we see that $|F_n^{(r)}|^2$ means the probability of a fermion with helicity $r$ propagates from in-vacuum towards out-vacuum, indicating a persistence probability to retain its helicity. 
This shows the same interpretation of the norm of effective action $|\exp\mathrm i S_n(k)|^2$, a vacuum-to-vacuum transition by tunnelling the scattering at $t_n$. 
Thus the norm of the eigenvalues of matrix $\exp\mathbf T$ are the correspondence of the diagonal elements $|F_n^{(r)}|^2$ of transfer matrix about Bogoliubov coefficients. 
On the other hand, off-diagonal elements $|H_n^{(r)}|^2$ represent the reflection probability of a fermion, which means a particle propagates to the asymptotic past and thus the helical state undergoes a flip. 
In this sense, for the Yukawa interaction with damping oscillation of inflaton \eqref{phi_t_damped}, after $N$ successive production events the occupation numbers of helical-symmetric fermions are given by 
\cite{Yukawa_preheating1,preheating_fermion1}
\begin{align}
    n^{(N)}(k)=\frac12-\frac12\prod_{j=1}^N\left[1-2\exp\left( -\frac{j\pi(k^{2}+m_{\psi}^{2})}{g\phi_{0}m_{\phi}} \right) \right]. 
\end{align}

The above analysis in this subsection is based on the “random-walk recipe,” in which the coherence between successive production events is neglected. 
To investigate the generation of helical fermions arising from oscillatory coherence, we now consider pseudoscalar fermionic preheating with a damped inflaton background. 
We use the total tunneling probabilities $\Gamma_N^{(\pm)}(k)$, given in Eq.~\eqref{Gamma_pm_k}, to characterize the production probabilities of each helical mode, assuming a vanishing initial fermion number. 
In this expression, $\lambda_{\pm}^{(N)}$ are the eigenvalues of the transfer matrix $\mathbf{R}^{(N)}$ defined in Eq.~\eqref{expR}. 
Figure~\ref{Gamma_Fig} shows the production probabilities $\Gamma_N^{(\pm)}(k)$ as functions of the dimensionless momentum $k/m_\psi$ after $N=1,3,5,8,25$ and $40$ successive scattering events, 
where the parameters are set $\omega_0^2=5\pi m_\psi^2$ and rescaled as $\mu_j=\omega_j^2$. . 
The results demonstrate that a net helical asymmetry is generated and survives after series of oscillations once a realistically damped inflaton background is taken into account. 
In the numerical calculations the diagonal elements $T_{11}$ and $T_{22}$ are regularized in the same manner as in Eq.~\eqref{Seff_regularization}, while the off-diagonal elements are regularized as
\begin{align}
    T_{12}^{(j)}=&\frac{\mathrm i}{4}(-1)^j\cdot\nonumber\\
    &\left[\int_{0}^{\infty}\frac{\mathrm{d}z}{z}\left(\mathrm{e}^{-\mathrm{i}(k^{2}+m_{\psi}^{2})z/\omega_j^2}-\mathrm{e}^{-\mathrm{i} \mu_j z/\omega_j^2}\right)-\frac{\mathrm i\pi}{2} \right].
\end{align}
The constant $\mathrm{i}\pi/2$ on the right-hand side comes from the integral
\begin{align}
    \operatorname{Im}\int_{0}^{\infty}\frac{\mathrm{d}z}{z}\mathrm{e}^{-\mathrm{i}(k^{2}+m_{\psi}^{2})z/\omega_j^2}=-\int_{0}^{\infty}\mathrm{d}z\frac{\sin z}{z}=-\frac\pi2. \label{T1122_damped_regu}
\end{align}
As will be seen, this constant plays an important role in the coherence between successive fermionic production events. 
For small $N$, a clear splitting between $\Gamma_N^{(+)}$ and $\Gamma_N^{(-)}$ appears. 
As the momentum $k$ increases, the cumulative production probabilities exhibit oscillatory behaviour and eventually damp to zero. 
For large $N$, the asymmetry becomes more pronounced at high momentum: both probabilities approach a step-like, saturated profile at small $k$ with a Gaussian-like profile, 
while they manifest as gradual saturation states at large $k$ controlled by the instantaneous $\omega_{j}$. 
The curves for $N=8$ and $N=40$ confirm that the helicity asymmetry persists and is concentrated at large $k$. 
In fact, an ultraviolet enhancement of both helicities occurs after every eight production events. 
This is because the retained phase $\pi/2$ in the off-diagonal elements $T_{12}^{(j)}$ and $T_{21}^{(j)}$, seen as in Eq.~\eqref{T1122_damped_regu}, 
accumulates to $2\pi$ after eight crossings, leading to a coherent amplification of fermion production, which is consistent with previous results \cite{Yukawa_preheating}. 
On the other hand, because the inflaton amplitude decreases ($\omega_{j}^{2}\propto 1/j$), each successive scattering possesses a narrower resonance window. 
This gradually suppresses the continued growth of the helicity asymmetry and eventually freeze it at a fixed value.

\section{Conclusions and Discussions \label{conclusion}}

In this work, we investigated fermionic preheating driven by a pseudoscalar (axion-like) Yukawa coupling, $g\phi\,\bar\psi\mathrm i\gamma^5\psi$, within the framework of the one-loop effective action. 
By employing the Schwinger proper time representation and projecting onto the helical basis, we derived the transfer matrix associated with each non-adiabatic crossing of the inflaton through zero. 
The diagonal elements of the transfer matrix encode the usual Gaussian pair production probability, while the off-diagonal elements capture the helicity-mixing amplitude that is characteristic of the pseudoscalar interaction. 

For an undamped oscillatory background the residual helicity mixing cancels after every even number of production events, so that a sustained net helical asymmetry cannot accumulate. 
Once a realistic damping of the inflaton amplitude is included, however, the exact time-reflection symmetry about each zero-crossing is broken. 
Consequently the off-diagonal contributions no longer cancel, and a net helical asymmetry is generated. 
Numerical evaluation of the accumulated transfer matrix after $N=1,3,5,8,25$ and $40$ oscillations shows that this asymmetry survives and becomes increasingly concentrated in the infrared as the resonance window narrows. 

We have also clarified the precise relation between the effective action transfer matrix $\exp\mathbf{T}$ and the conventional Bogoliubov coefficients: 
the eigenvalues of the former directly determines the helical occupation numbers, while the off-diagonal entries correspond to the helicity-flip amplitudes. 
This establishes a transparent bridge between the effective action approach and the more traditional mode function analyses. 

Several extensions may be promised in this framework. 
First, a fully expanding-universe treatment that incorporates both Hubble damping and the continuous redshift of physical momenta would allow a quantitative prediction of the final comoving helical spectrum. 
Second, the back-reaction of the produced fermions on the inflaton condensate, as well as the possible generation of helical gravitational waves or primordial magnetic fields, deserves a dedicated study. 
Third, the inclusion of thermal effects and chemical potentials could significantly modify the fluctuation-dissipation relation and the resulting asymmetry, especially relevant if the preheating stage transitions into a warm-inflationary regime. 
Finally, the same effective action technology can be applied to mixed scalar and pseudoscalar couplings or to higher-dimensional operators, 
opening a potential method to explore helicity-asymmetric particle production in a broad class of early universe models. 

In summary, the combination of a pseudoscalar Yukawa interaction with a realistically damped inflaton background provides a natural and efficient mechanism for generating a net fermionic helicity asymmetry during preheating. 
The effective-action approach developed here offers both analytical insight and a practical computational framework for further investigations of this phenomenon and its cosmological consequences.

\begin{acknowledgements}
This work was supported by the National Natural Science Foundation of China (Grant No. 12275143), Central Guidance for Local Science and Technology Development Fund Project (Grands No. 2024ZY0113, 2025ZY0020),
Inner Mongolia Natural Science Foundation (Grants No. 2024SHZR0009, 2026LHMS0083, 2026LHQC0206).
\end{acknowledgements}


\begin{thebibliography}{999}

\bibitem{reheating1} B. A. Bassett, S. Tsujikawa, and D. Wands, Inflation dynamics and reheating, Rev. Mod. Phys. \textbf{78}, 537 (2006).

\bibitem{preheating_boson01} Y. Shtanov, J. H. Traschen, and R. H. Brandenberger, Universe reheating after inflation, Phys. Rev. D \textbf{51}, 5438 (1995).

\bibitem{preheating1} M. A. Amin, M. P. Hertzberg, D. I. Kaiser, and J. Karouby, Nonperturbative dynamics of reheating after inflation: A review, Int. J. Mod. Phys. D \textbf{24}, 1530003 (2015).

\bibitem{preheating_boson02} J. H. Traschen and R. H. Brandenberger, Particle production during out-of-equilibrium phase transitions, Phys. Rev. D \textbf{42}, 2491 (1990).

\bibitem{preheating_boson1} L. Kofman, A. D. Linde, and A. A. Starobinsky, Reheating after inflation, Phys. Rev. Lett. \textbf{56}, 3258 (1997).

\bibitem{preheating_boson2} L. Kofman, A. D. Linde, and A. A. Starobinsky, Towards the theory of reheating after inflation, Phys.D \textbf{56}, 3258 (1997). 

\bibitem{preheating_boson3} F. Pineda and L. O. Pimentel, Non-perturbative approach for scalar particle production in Higgs-$R^2$ inflation, Eur. Phys. J. C \textbf{85}, 731 (2025).

\bibitem{preheating_boson4} S. S. Mishra and V. Sahni, New models of quintessential inflation featuring plateau and hilltop potentials, Eur. Phys. J. C \textbf{85}, 48 (2025).

\bibitem{effective_boson_preheating} B. Xu and W. Xue, Effective action approach for preheating, J. Cosmol. Astropart. Phys. 05 (2024) 038.

\bibitem{preheating_fermion1} M. Peloso and L. Sorbo, Preheating of massive fermions after inflation: Analytical results, J. High Energy Phys. 05 (2000) 016.

\bibitem{preheating_fermion2} L. Sorbo, Fermion preheating: Analytical results, Nucl. Phys. B Proc. Suppl. \textbf{95}, 86 (2001).

\bibitem{preheating_fermion3} J. García-Bellido, S. Mollerach, and E. Roulet, Fermion production during preheating after hybrid inflation, J. High Energy Phys.  02 (2000) 034.

\bibitem{preheating_fermion4} G. F. Giudice, A. Riotto, I. Tkachev, and M. Peloso, Production of massive fermions at preheating and leptogenesis, J. High Energy Phys. 08 (1999) 014.

\bibitem{preheating_fermion5} P. B. Greene and L. Kofman, Preheating of fermions, Phys. Lett. B  \textbf{448}, 6 (1999).

\bibitem{preheating_fermion6} S. Tsujikawa, B. A. Bassett, and F. Viniegra, Multifield fermionic preheating, J. High Energy Phys. 08 (2000) 019.

\bibitem{preheating_fermion7} H. E. Logan, D. Stolarski, and F. Yasin, Analytic approximations for fermionic preheating, Phys. Rev. D \textbf{114}, 076012 (2026).

\bibitem{preheating_fermion8} J. Klaric, A. Shkerin, and G. Vacalis, Non-perturbative production of fermionic dark matter from fast preheating, J. Cosmol. Astropart. Phys. 02 (2023) 034.

\bibitem{Yukawa_preheating} K. Fu, L.-S. Liu, Y.-F. Wang, X.-Y. Gu, and X.-B. Li, Pseudoscalar Yukawa coupling induced helical asymmetry in fermionic preheating, Eur. Phys. J. C \textbf{86}, 1083 (2026).

\bibitem{Yukawa_preheating1} P. Adshead and E. I. Sfakianakis, Fermion production during and after axion inflation, J. Cosmol. Astropart. Phys. 11 (2015) 021.

\bibitem{QCD1} X.-R. Wang, S. Matsuzaki, J.-Y. Li, S. Enomoto, and H. Ishida, QCD preheating: New frontier of baryogenesis, Phys. Rev. D \textbf{108}, 023512 (2023).

\bibitem{QCD2} J. Wang, X.-R. Wang, and S. Matsuzaki, Baryogenesis via QCD preheating with nonadiabatic baryon chemical potential, J. High Energy Phys. 08 (2024) 032.

\bibitem{preheating_GW1} S. Y. Khlebnikov and I. I. Tkachev, Relic gravitational waves produced after preheating, Phys. Rev. D \textbf{56}, 653 (1997).

\bibitem{preheating_GW2} R. Easther, J. T. Giblin, Jr., and E. A. Lim, Gravitational wave production at the end of inflation, Phys. Rev. Lett. \textbf{99}, 221301 (2007).

\bibitem{preheating_GW3} P. Adshead, J. T. Giblin, Jr., and A. Tishue, Gravitational waves from kinetic preheating, Phys. Rev. D \textbf{110}, 043536 (2024).

\bibitem{preheating_GW4} J. García-Bellido, D. G. Figueroa, and A. Sastre, A gravitational wave background from reheating after hybrid inflation, Phys. Rev. D \textbf{77}, 043517 (2008).

\bibitem{preheating_GW5} J.-F. Dufaux, A. Bergman, G. Felder, L. Kofman, and J.-P. Uzan, Theory and numerics of gravitational waves from preheating after inflation, Phys. Rev. D \textbf{76}, 123517 (2007).

\bibitem{preheating_GW6} J.-F. Dufaux, D. G. Figueroa, and J. García-Bellido, Gravitational waves from Abelian gauge fields and cosmic strings at preheating, Phys. Rev. D \textbf{82}, 083518 (2010). 

\bibitem{preheating_GW7} J. Kim, Z. Yang, and Y.-L. Zhang, Gravitational wave signatures of preheating in Higgs-R2 inflation, Phys. Rev. D \textbf{112}, 043534 (2025).

\bibitem{preheating_GW8} L. Bethke, D. G. Figueroa, and A. Rajantie, On the anisotropy of the gravitational wave background from massless preheating, J. Cosmol. Astropart. Phys. 06 (2014) 047.

\bibitem{preheating_GW9} P. Adshead, J. T. Giblin, Jr., and Z. J. Weiner, Gravitational waves from gauge preheating, Phys. Rev. D \textbf{98}, 043525 (2018).

\bibitem{preheating_GW10} C. Cosme, D. G. Figueroa, and N. Loayza, Gravitational wave production from preheating with trilinear interactions, J. Cosmol. Astropart. Phys. 05 (2023) 023.

\bibitem{preheating_GW11} W.-Y. Hu, K. Nakayama, V. Takhistov, and Y. Tang, Dual gravitational wave signatures of instant preheating, J. Cosmol. Astropart. Phys. 01 (2025) 029.

\bibitem{preheating_GW12} D. G. Figueroa and F. Torrenti, Gravitational wave production from preheating: Parameter dependence, J. Cosmol. Astropart. Phys. 10 (2017) 057.

\bibitem{preheating_GW13} C. Caprini and D. G. Figueroa, Cosmological backgrounds of gravitational waves, Class. Quantum Grav. \textbf{35},  163001 (2018).

\bibitem{preheating_GW14} N. Bartolo, et al, Probing non-Gaussian stochastic gravitational wave backgrounds with LISA, J. Cosmol. Astropart. Phys. 11 (2018) 034.

\bibitem{preheating_GW15} K. D. Lozanov and M. A. Amin, Gravitational perturbations from oscillons and transients after inflation, Phys. Rev. D \textbf{99}, 123504 (2019).

\bibitem{preheating_GW16} P. Adshead, J. T. Giblin, Jr., M. Pieroni, and Z. J. Weiner, Constraining axion inflation with gravitational waves from preheating, Phys. Rev. D \textbf{101}, 083534 (2020).

\bibitem{preheating_NG1} K. Enqvist, et al, Non-Gaussianity from preheating, Phys. Rev. Lett. \textbf{94}, 161301 (2005).

\bibitem{preheating_NG2} K. Enqvist, A. Jokinen, A. Mazumdar, T. Multamäki, and A. Väihkönen, Non-Gaussianity from instant and tachyonic preheating, J. Cosmol. Astropart. Phys. 03 (2005) 010.

\bibitem{preheating_NG3} A. Jokinen and A. Mazumdar, Very large primordial non-Gaussianity from multi-field: Application to massless preheating, J. Cosmol. Astropart. Phys. 04 (2006) 003.

\bibitem{preheating_NG4} N. Barnaby and J. M. Cline, Nongaussian and nonscale-invariant perturbations from tachyonic preheating in hybrid inflation, Phys. Rev. D \textbf{73}, 106012 (2006).

\bibitem{preheating_NG5} N. Barnaby and J. M. Cline, Nongaussianity from tachyonic preheating in hybrid inflation, Phys. Rev. D \textbf{75}, 086004 (2007).

\bibitem{preheating_NG6} K. Kohri, D. H. Lyth, and C. A. Valenzuela-Toledo, Preheating and the non-Gaussianity of the curvature perturbation, J. Cosmol. Astropart. Phys. 02 (2010) 023.

\bibitem{preheating_NG7} J. R. Bond, A. V. Frolov, Z. Huang, and L. Kofman, Non-Gaussian spikes from chaotic billiards in inflation preheating, Phys. Rev. Lett. \textbf{103}, 071301 (2009).

\bibitem{preheating_NG8} G. Leung, E. R. M. Tarrant, C. T. Byrnes, and E. J. Copeland, Reheating, multifield inflation and the fate of the primordial observables, J. Cosmol. Astropart. Phys. 09 (2012) 008.

\bibitem{preheating_NG9} G. Leung, E. R. M. Tarrant, C. T. Byrnes, and E. J. Copeland, Influence of reheating on the trispectrum and its scale dependence, J. Cosmol. Astropart. Phys. 08 (2013) 006.

\bibitem{preheating_NG10} S. V. Imrith, D. J. Mulryne, and A. Rajantie, Primordial curvature perturbation from lattice simulations, Phys. Rev. D \textbf{100}, 043543 (2019).

\bibitem{preheating_NG11} J. Fan and Z.-Z. Xianyu, A cosmic microscope for the preheating era, J. High Energy Phys. 01 (2021) 021.

\bibitem{preheating_NG12} R. Grutkoski, H. J. Macpherson, J. T. Giblin, Jr., and J. Frieman, Effect of nonlinear gravity on the cosmological background during preheating, Phys. Rev. D \textbf{112}, 023541 (2025).

\bibitem{preheating_PB1} E. A. Calzetta and A. Kandus, Selfconsistent estimates of magnetic fields from reheating, Phys. Rev. D \textbf{65}, 063004 (2002).

\bibitem{preheating_PB2} A. C. Davis, K. Dimopoulos, T. Prokopec, and O. Törnkvist, Primordial spectrum of gauge fields from inflation, Phys. Lett. B \textbf{501}, 165 (2001).

\bibitem{preheating_PB3} D. Boyanovsky, H. J. de Vega, and M. Simionato, Large scale magnetogenesis from a nonequilibrium phase transition in the radiation dominated era, Phys. Rev. D \textbf{67}, 123505 (2003).

\bibitem{preheating_PB4} D. Boyanovsky, M. Simionato, and H. J. de Vega, Magnetic field generation from nonequilibrium phase transitions, Phys. Rev. D \textbf{67}, 023502 (2003).

\bibitem{preheating_PB5} A. Mazumdar and H. Stoica, Exciting gauge field and gravitons in a brane-antibrane annihilation, Phys. Rev. Lett. \textbf{102}, 091601 (2009).

\bibitem{preheating_PB6} A. Díaz-Gil, J. García-Bellido, M. García Pérez, and A. González-Arroyo, Magnetic field production during preheating at the electroweak scale, Phys. Rev. Lett. \textbf{100}, 241301 (2008).

\bibitem{preheating_PB7} A. Bhaumik, T. Papanikolaou, and A. Ghoshal, Vector induced gravitational waves sourced by primordial magnetic fields, J. Cosmol. Astropart. Phys. 08 (2025) 054.

\bibitem{preheating_PB8} L. C. Garcia de Andrade, Generation of primordial magnetic fields from QED and Higgs-like domain walls in Einstein–Cartan gravity, Universe \textbf{8}, 658 (2022).

\bibitem{preheating_BH1} A. M. Green and K. A. Malik, Primordial black hole production due to preheating, Phys. Rev. D \textbf{64}, 021301(R) (2001).

\bibitem{preheating_BH2} B. A. Bassett and S. Tsujikawa, Inflationary preheating and primordial black holes, Phys. Rev. D \textbf{63}, 123503 (2001).

\bibitem{preheating_BH3} T. Suyama, T. Tanaka, B. Bassett, and H. Kudoh, Are black holes over-produced during preheating?, Phys. Rev. D \textbf{71}, 063507 (2005).

\bibitem{preheating_BH4} M. R. Haque, E. Kpatcha, D. Maity, and Y. Mambrini, Primordial black hole reheating, Phys. Rev. D \textbf{108}, 063523 (2023).

\bibitem{preheating_BH5} L. E. Padilla, J. C. Hidalgo, T. D. Gómez-Aguilar, K. A. Malik, and G. Germán, Primordial black hole formation during slow-reheating: A review, Front. Astron. Space Sci. \textbf{11}, 1361399 (2024).

\bibitem{preheating_BH6} D. del-Corral, P. Gondolo, K. Sravan Kumar, and J. Marto, Primordial black holes through preheating instabilities in $\alpha$-attractor models, Phys. Rev. D 114, 023561 (2026).

\bibitem{preheating_BH7} D. del-Corral, P. Gondolo, K. Sravan Kumar, and J. Marto, Revisiting primordial black holes formation from preheating instabilities: The case of Starobinsky inflation, J. Cosmol. Astropart. Phys. 02 (2025) 009.

\bibitem{preheating_BH8} P. Adshead, E. Currens, and J. T. Giblin, Jr., Primordial black holes from kinetic preheating, Phys. Rev. D \textbf{114}, L021304 (2026).

\bibitem{preheating_BH9} G. Ballesteros, J. Iguaz Juan, P. D. Serpico, and M. Taoso, Primordial black hole formation from self-resonant preheating?, Phys. Rev. D \textbf{111}, 083521 (2025).

\bibitem{preheating_TD1} S. Khlebnikov, L. Kofman, A. D. Linde, and I. Tkachev, First order nonthermal phase transition after preheating, Phys. Rev. Lett. \textbf{81}, 2012 (1998).

\bibitem{preheating_TD2} I. Tkachev, S. Khlebnikov, L. Kofman, and A. D. Linde, Cosmic strings from preheating, Phys. Lett. B \textbf{440}, 262 (1998).

\bibitem{preheating_TD3} M. F. Parry and A. T. Sornborger, Domain wall production during inflationary reheating, Phys. Rev. D \textbf{60}, 103504 (1999).

\bibitem{preheating_TD4} J. Lee, K. Murai, K. Nakayama, and F. Takahashi, Dynamical prevention of topological defect formation, J. High Energy Phys. 03 (2026) 005.

\bibitem{GUT1} G. F. Giudice, A. Riotto, I. Tkachev, and M. Peloso, Production of massive fermions at preheating and leptogenesis, J. High Energy Phys. 08 (1999) 014.

\bibitem{GUT2} R. Barbieri, P. Creminelli, A. Strumia, and N. Tetradis, Baryogenesis through leptogenesis, Nucl. Phys. B \textbf{575}, 61 (2000).

\bibitem{GUT3} A. Datta, R. Roshan, and A. Sil, Flavor leptogenesis during the reheating era, Phys. Rev. D \textbf{108}, 035029 (2023).

\bibitem{bary1} C. Chen, S. J. Das, K. Dimopoulos, and A. Ghoshal, Flipped rotating axion non-minimally coupled to gravity: Baryogenesis and dark matter, Eur. Phys. J. C \textbf{85}, 898 (2025).

\bibitem{WKB1} M. V. Berry and K. E. Mount, Semiclassical approximations in wave mechanics, Rep. Prog. Phys. \textbf{35}, 315 (1972).

\bibitem{WKB2} E. Brézin and C. Itzykson, Pair production in vacuum by an alternating field, Phys. Rev. D \textbf{2}, 1191 (1970).

\bibitem{WKB3} J. Schwinger, On gauge invariance and vacuum polarization, Phys. Rev. \textbf{82}, 664 (1951).

\bibitem{pseudoscalar1} A. Berlin, S. Gori, T. Lin, and L. T. Wang, Pseudoscalar portal dark matter, Phys. Rev. D \textbf{92}, 015005 (2015).

\bibitem{pseudoscalar2} S. Biondini, J. Bollig, and S. Vogl, Indirect detection of dark matter with (pseudo)-scalar interactions, J. High Energy Phys.  \textbf{04} (2024) 050.

\bibitem{Schwinger_time1} T. Padmanabhan, Quantum theory in external electromagnetic and gravitational fields: A comparison of some conceptual issues, Pramana \textbf{37}, 179 (1991).

\bibitem{Schwinger_time2} M. D. Schwartz, Quantum Field Theory and the Standard Model (Cambridge University Press, Cambridge, 2013).

\bibitem{instanton} S. Coleman, Aspects of Symmetry: Selected Erice Lectures (Cambridge University Press, Cambridge, 1985).

\end{thebibliography}
\end{document}